\documentstyle[12pt,epsf]{article}
\begin{document}
\begin{flushright}
KEK-CP-102 \\
Nov. 09, 2000
\end{flushright}
\vskip 2cm
\begin{center}
{\Large
QED Radiative Correction for the Single-$W$ Production
using a Parton Shower Method
}
\end{center}
\vskip 1cm
\begin{center}
{\large Y. Kurihara, J. Fujimoto, T. Ishikawa, Y. Shimizu} \\
{\it High Energy Accelerator Research Organization,\\
Tsukuba, 305-0801, Japan}\\
\vskip 5mm
{\large K. Kato, K. Tobimatsu} \\
{\it Kogakuin University, Shinjuku, Tokyo 163-8677, Japan} \\
\vskip 5mm
{\large T. Munehisa} \\
{\it Yamanashi University, Yamanashi 400-8510, Japan}
\end{center}
\vskip 1cm
\begin{abstract}
A parton shower method for the photonic radiative correction 
is applied to the single $W$-boson production processes.
The energy scale for the evolution of the parton shower is determined
so that the correct soft-photon emission is reproduced. Photon spectra 
radiated from the partons are compared with those from the exact matrix
elements, and show a good agreement. Possible errors due to a inappropriate
energy-scale selection or due to the ambiguity of energy scale 
determination are also discussed,
particularly for the measurements on triple gauge-couplings.
\end{abstract}
\newpage

\section{Introduction}
Single-$W$ production processes present
an opportunity to study the anomalous triple gauge-couplings(TGC)\cite{tuka}
in the experiments at LEP2 and at future $e^+e^-$ linear colliders. 
For their precise predictions of cross sections
the inclusion of an initial state 
radiative correction(ISR) is inevitable in the event-generator. 
As a tool for the ISR the structure function(SF)\cite{grace:SF}
and the parton shower\cite{grace:PS}
methods has been widely used for $e^+ e^-$ annihilation processes.
For the case of the single-$W$ production processes, however,
the main contribution comes from non-annihilation type diagrams.
The universal factorization method used for the annihilation processes 
is obviously inappropriate. The main problem lies in choosing 
the energy scale of the factorization. A previous
study on the two photon process\cite{grace:twopho} has shown that
the SF and QED parton shower(QEDPS) methods are able to
reproduce precisely the exact $O(\alpha)$ results 
even for non-annihilation processes, as long as the appropriate energy 
scale is used.

In this report a general method to find the energy scale for SF and QEDPS 
is proposed. Then numerical results of testing SF and QEDPS for 
$e^- e^+ \rightarrow e^- {\bar \nu}_e u {\bar d}$ and
$e^- e^+ \rightarrow e^- {\bar \nu}_e \mu^+ \nu_\mu$
are presented. Systematic errors are also discussed.
 
\section{Calculation method}
\subsection{Energy Scale Determination}
The factorization theorem for QED radiative corrections in 
the leading-logarithmic approximation is valid independent of 
the structure of the matrix element of the kernel process. 
Hence SF and QEDPS must be applicable to {\it any} $e^+e^-$ scattering 
processes. However, choice of the energy scale in SF and QEDPS is not 
a trivial matter\cite{bbn}.
For simple processes considered so far like $e^+e^-$ annihilation and 
two-photon process with only multi-peripheral diagrams, the evolution 
energy scale could be found by the exact perturbative calculations. 
However, this is not always the case when more complicated processes
are concerned. Hence a way to find a suitable energy scale without 
knowing exact loop calculations should be established somehow.

First let us look at the general consequence of the soft photon
approximation. The cross section with radiations in the soft-photon limit 
is given by the Born cross section multiplied by the following factor 
up to the double-log term as
\cite{grace:SW}
\begin{eqnarray}
\frac{d\sigma_{soft}(s)}{d\Omega} &=& 
\frac{d\sigma_0(s)}{d\Omega} \nonumber \\
&\times& \left| {\exp}\left[-\frac{\alpha}
{\pi} {\ln}\left( \frac{E}{k_c} \right)  
\sum_{i,j} \frac{e_i e_j \eta_i \eta_j}
{\beta_{ij}} {\ln}\left(\frac{1+\beta_{ij}}{1-\beta_{ij}}\right) 
\right] \right|^2,
\label{ir} \\
\beta_{ij} &=& \left(1-\frac{m_i^2 m_j^2}
{(p_i \cdot p_j)^2}\right)^{\frac{1}{2}},
\end{eqnarray}
where $m_j$($p_j$) are the mass(momentum) of the $j$-th charged 
particle, $k_c$ the maximum energy of the soft photon(the value to separate
soft- and hard-photons), $E$ the beam energy and $e_j$ the electric 
charge in unit of the $e^+$ charge. The factor $\eta_j$ is $-1$ for 
the initial particles and $+1$ for the final particles. The indices 
($i,j$) run over all the charged particles in the initial and final 
states. 

For the two-photon process,
$e^-(p_-) e^+(p_+) \rightarrow e^-(q_-) e^+(q_+) \mu^-(k_-) \mu^+(k_+)$,
it is shown in ref.\cite{grace:twopho} 
that the soft-photon factor in Eq.(\ref{ir}) with a ($p_- \cdot q_-$)-term 
reproduces the $O(\alpha)$ corrections\cite{grace:bdk} 
up to the double-log term in the soft-photon limit. This implies that 
one is able to read off the possible evolution energy scale in SF from 
Eq.(\ref{ir}) without doing explicit loop calculations.\footnote{
A similar idea is independently proposed by Montagna et al.~in\cite{mont}.}
The point is the observation that the energy scale $s=(p_-+p_+)^2$ 
does not appear in the soft-photon corrections even they are included 
in the general formula Eq.(\ref{ir}). In the case of the two-photon 
process we have ignored those terms in SF which come from the photon 
bridged between different charged lines, because the contributions
from the box diagrams with a photon exchange between $e^+$ and $e^-$ is 
known to be small\cite{grace:vNV}. Fortunately the infrared part of 
the loop corrections is already included in Eq.(\ref{ir}) and there is
no need to know the full form of the loop diagram. Let us look at two 
terms with, for example ($p_- \cdot p_+$)- and ($q_- \cdot p_+$)-terms.
The momentum of $e^-$ is almost 
the same before and after the scattering($p_- \approx q_-$). 
Only the difference appears in $\eta_j \eta_k=+1$ for a ($p_-  p_+$)-term  
and $\eta_j \eta_k=-1$ for a ($q_- p_+$)-term. Then these terms compensate 
each other after summing them up for the forward scattering which is 
the dominant kinematical region of this process. This is the mechanism
that the energy scale $s=(p_-+p_+)^2$ does not appear in the soft-photon 
correction despite that it exists in Eq.(\ref{ir}).

When some experimental cuts are imposed, for example the final $e^-$ is
tagged in a large angle, this cancellation is not perfect but partial 
and the energy scale $s$ must appear in the soft-photon correction. 
In this case the annihilation type diagrams will also give a 
contribution. Then it may happen that the usual SF and QEDPS for the
annihilation processes are justified to be used for the ISR with 
the energy scale $s$. To find the most dominant energy scale under 
the given experimental cuts an easy way is to integrate numerically
the soft-photon cross section given by Eq.(\ref{ir}) over the allowed
kinematical region. Thus in order to determine the energy scale it is 
sufficient to know the infrared behavior of the radiative process using 
the soft-photon factor. 
%In some region of the phase-space two or more 
%energy scales may be involved in the soft-photon cross section with
%comparable amount of contribution. In this region a simple the SF and 
%the QEDPS are not applicable.

Let us determine the energy-scale of the QED radiative corrections
to the single-$W$ production process,
\begin{eqnarray}
e^-(p_-)+e^+(p_+) &\rightarrow& e^-(q_-)+{\bar \nu}_e(q_{\nu})
+u(k_u)+{\bar d}(k_d).
\label{pc1}
\end{eqnarray}
The soft-photon correction factor in 
Eq.(\ref{ir}) is numerically integrated with the Born matrix element of
the process (\ref{pc1}) only with the $t$-channel diagrams without any 
cut on the final fermions. 
In order to separate the contribution from each term,
we take terms up to $O(\alpha)$ in Taylor expansion
of exponential function in Eq.(\ref{ir}). 
Parameters used in the calculation will be explained
in section 4. 
The results are shown in Table \ref{grace:TB1}.
One can see that the main contribution 
comes from an electron charged-line ($p_- q_-$-term) and 
a positron charged-line 
($p_+ k_u k_d $-terms=$p_+ k_u$-term + $p_+ k_d$-term + $k_u k_d $-term), 
while all other contributions are negligiblly small.
Like the two-photon processes the energy scale 
$s$ does not appear in the soft-photon correction.
%%%%%%%%%%%%%%%%%%%%%%%%%% table 1 %%%%%%%%%%%%%%%%%%%%%%%%%%%%%%%%%%%%%%%
\begin{table}[htbp]
\begin{center}
\begin{tabular}{|c|c|c|c|} \hline
all terms & $p_- q_-$ & $p_+ k_u k_d $  & all other combinations \\ \hline
    1     &  $0.38$     &  $0.61$     & $1.9\times10^{-3}$    \\ \hline
\end{tabular}
\end{center}
\caption{\footnotesize
Soft-photon correction factor from sets of the charged particle
combinations for the process of $e^+ e^- \rightarrow e^- {\bar \nu}_e 
u {\bar d}$ at the CM energy of 200 GeV. The factor from all terms
is normalized to unity. $k_c=1$GeV is used.
$p_+ k_u k_d $=$p_+ k_u$-term + $p_+ k_d$-term + $k_u k_d $-term.
}
\label{grace:TB1}
\end{table}
%%%%%%%%%%%%%%%%%%%%%%%%%% table 1 %%%%%%%%%%%%%%%%%%%%%%%%%%%%%%%%%%%%%%%
The above results clearly indicate that one should apply SF or QEDPS
to the electron and positron charged-lines individually with the energy 
scale of their momentum-transfer squared.

\subsection{Structure Function Method}
The analytic solution of the DGLAP evolution function\cite{dglap}
in the leading-logarithmic order is known as the structure 
function\cite{grace:SF}.
With SF the QED corrected cross section is given by
\begin{eqnarray}
\sigma_{total}(s) &=&\int dx_{I-}\int dx_{F-}
\int dx_{I+} \int dx_{u} \int dx_{d}\nonumber 
D_{e^-}(x_{I-},-t_-) D_{e^-}(x_{F-},-t_-) \\
&~&D_{e^+}(x_{I+},-t_+) D_{u}(x_{u},s_{ud}) D_{d}(x_{d},s_{ud}) 
\sigma_0({\hat s}), 
\label{correction}
\end{eqnarray}
where $\sigma_0$ is the Born cross section and $D$'s are the SF.
The energy scales $t_-=(p_- - q_-)^2$, $t_+=(p_+ - (k_u+k_d))^2$
and $s_{ud}=(k_u+k_d)^2$ are chosen following the result of 
the last section. 
%The energy-scale determination in each SF on the positron side is rather
%ambiguous. The $p_{T u+d}$ is distributing around $M_W/3$, then
%the difference between these two energy scales does not give
%a significant effect on the correction factor.
After(before) the photon radiation the initial(final) momenta 
$p_\pm$($q_\pm$) become ${\hat p_\pm}$ (${\hat q_\pm}$) 
\begin{eqnarray}
{\hat p_-} &=& x_{I-}p_-,~~~{\hat q_-}=\frac{1}{x_{F-}}q_-, ...
\end{eqnarray}
respectively. Then the CM energy squared, $s$, is scaled as 
${\hat s} = x_{I-}x_{I+}s$.

\subsection{Parton Shower Method}
Instead of the analytic formula of SF a Monte Calro method based on 
the parton shower algorithm in QED can be used to solve DGLAP equation 
in the LL approximation. Its detailed algorithm of the QEDPS is found 
in ref.\cite{grace:QEDPSs} for the $e^+e^-$ annihilation processes, 
in ref.\cite{grace:QEDPSt} for the Bhabha process, and in 
ref.\cite{grace:twopho} for the two-photon process. The same energy scale 
as the SF method is used in QEDPS also.
%One difference between SF and QEDPS is that 
%the {\it ad hoc} replacement of the perturbative expansion
%coefficient $L(={\ln}({Q^2}/{m_f^2}))$ by $L-1$, which was
%realized by hand for the SF, cannot be done for the QEDPS. 
Significant difference 
between SF and QEDPS is that the QEDPS can treat the transverse momentum 
of the emitted photons correctly by imposing the exact kinematics at the
$e\rightarrow e\gamma$ splitting. It does not affect the total cross 
sections so much when the final $e^-$ has no cut. However, the finite 
recoiling of the final $e^\pm$ can result some effects on the tagged 
cross sections.

In return for the exact kinematics at the $e\rightarrow e\gamma$ splitting, 
$e^\pm$ are no more on-shell after a photon emission. On the other hand 
the matrix element of the hard scattering process must be calculated with 
on-shell external particles. A trick to map the off-shell four-momenta of 
the initial $e^\pm$ to those at on-shell is needed. The following method 
is used in the calculations.
\begin{enumerate}
\item ${\hat s}=({\hat p}_- + {\hat p}_+)^2$ is calculated, where
${\hat p}_\pm$ are the four-momenta of the initial $e^\pm$ after the
photon emission by QEDPS. ${\hat s}$ is positive even for the 
off-shell $e^\pm$. 
\item 
New four-momenta
of initial $e^{\pm}$ in their rest-frame, ${\tilde p}_\pm$, are defined as
${\tilde p}_\pm^2=m_e^2$ (on-shell) 
and ${\hat s}=({\tilde p}_- + {\tilde p}_+)^2$.
All four-momenta of final particles 
are generated in the rest-frame of ${\tilde p}_+ + {\tilde p}_-$.
\item All four-momenta are rotated and boosted to match the
three-momenta of ${\tilde p}_\pm$ with those of ${\hat p}_\pm$.
\end{enumerate}
This method respects the direction of the final $e^\pm$ rather than 
the CM energy of the collision. The total energy does not conserve then
because of the virtuarity of the initial $e^\pm$.
The violation of the energy conservation is of the order of $10^{-6}$ GeV
or less, and the probability of the violation more than 1 MeV is $10^{-4}$.

\section{Numerical Calculations}

\subsection{Cross sections with no cut}
\begin{figure}[htb]
\centerline{
\epsfysize=15cm
\epsfbox{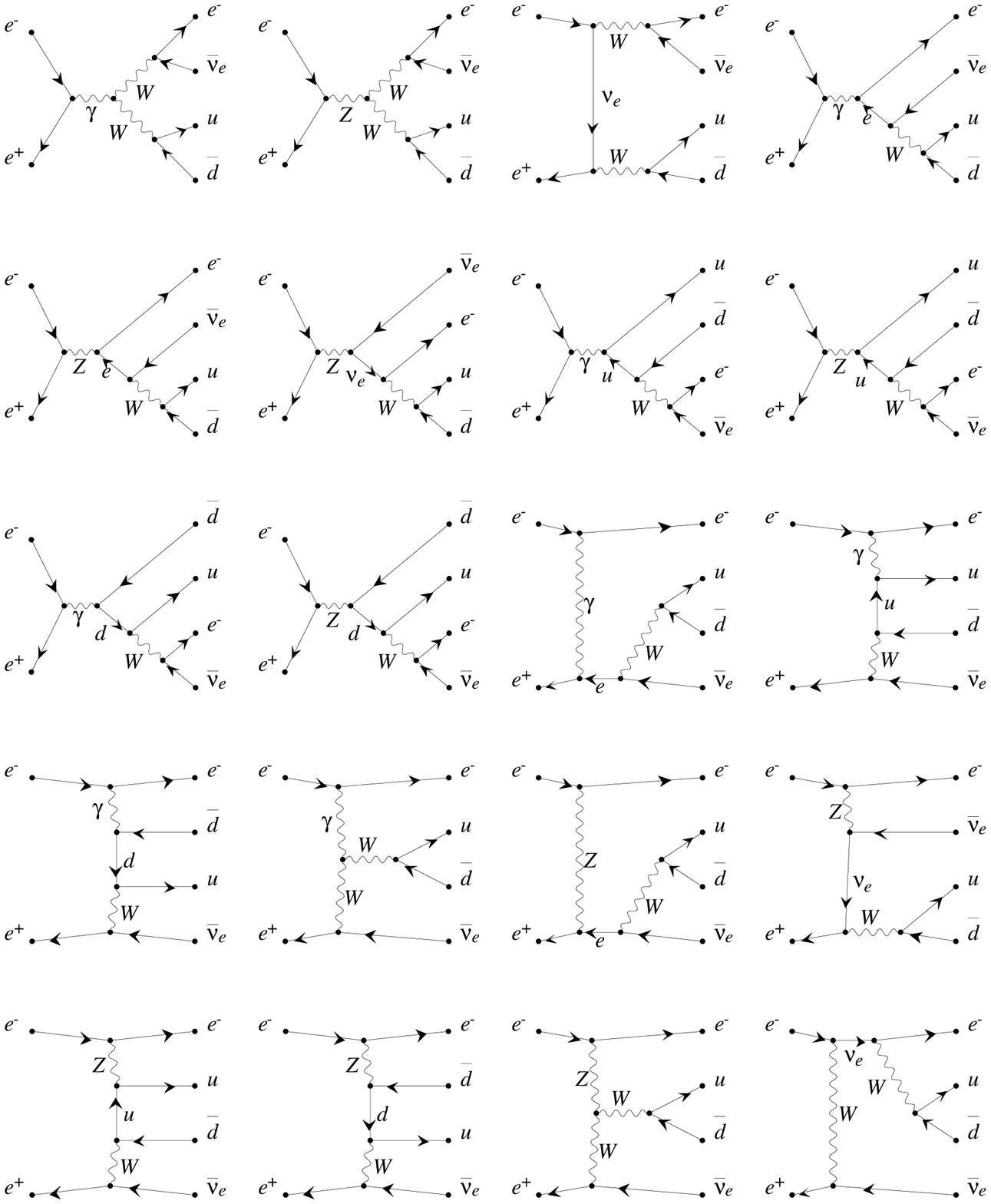}
}
\caption{
Feynman diagrams of the process 
$e^- e^+ \rightarrow e^- {\bar \nu}_e u {\bar d}$.
First ten diagrams show the $s$-channel diagrams and last ten 
the $t$-channel ones.
}
\label{fig-grace-diag}.
\end{figure}

Total and differential cross sections of semi-leptonic process 
$e^- e^+ \rightarrow e^- {\bar \nu}_e u {\bar d}$ and leptonic one 
$e^- e^+ \rightarrow e^- {\bar \nu}_e \mu^+ \nu_\mu$,
are calculated with the radiative correction by SF or QEDPS.
Feynman diagrams of the semi-leptonic process are shown 
in Fig.\ref{fig-grace-diag}. Fortran codes to calculate the amplitudes 
are automatically produced by ${\tt GRACE}$ system\cite{grace:grace}. 
All fermion-masses are kept finite in the calculations.
Numerical integrations of the matrix element squared in the four-body
phase space are done using ${\tt BASES}$\cite{grace:bases}.
For a test with no experimental cuts,
only the $t$-channel diagrams(non-annihilation diagrams) are taken into 
account. Standard model parameters used in the calculations are 
summarized in Table\ref{grace:TB2}. The on-shell relation strict at 
the tree-level is employed to determine weak couplings. 
Some of electro-weak corrections
could be taken into account through the $G_\mu$-scheme and running coupling
constant. In this report, however, we do not include those effects
because here we are interested in looking at the pure QED radiative 
corrections to the Born cross section. 
%%%%%%%%%%%%%%%%%%%%%%
\begin{table}[htbp]
\begin{center}
\begin{tabular}{c l c l}
\hline
$M_W$&80.35 GeV  &$\Gamma_W$&1.96708 GeV \\ 
$M_Z$&91.1867 GeV&$\Gamma_Z$&2.49471 GeV \\ 
$\alpha$&1/137.0359895& $\sin^2{\theta_w}$&$1-\frac{M_W^2}{M_Z^2}$ \\ 
$m_e$&$0.511 \times 10^{-3}$ GeV& $m_{\mu}$&$105.658389 \times 10^{-3}$ GeV
\\ 
$m_u$&$5.0 \times 10^{-3}$ GeV &
$m_d$&$10.0 \times 10^{-3}$ GeV \\ 
\hline
\end{tabular}
\end{center}
\caption{
Standard-model parameters
}
\label{grace:TB2}
\end{table}
%%%%%%%%%%%%%%%%%%%%%%
It has been pointed out that a tiny violation of the gauge invariance 
caused by the inclusion of a finite $W$-boson width results a wrong cross 
section for single-$W$ production\cite{kur,hlf}.
To cure this problem the fermion-loop scheme\cite{hlf,aco,FL} has been
proposed. It is reported in ref.\cite{aco}, however, that no significant 
difference is seen between the fermion-loop scheme and the fixed width 
scheme. Thus in this report the fixed width scheme, together with an
appropriately modified current, described in \cite{kur} is employed.

\begin{figure}[htb]
\centerline{
\epsfysize=15cm
\epsfbox{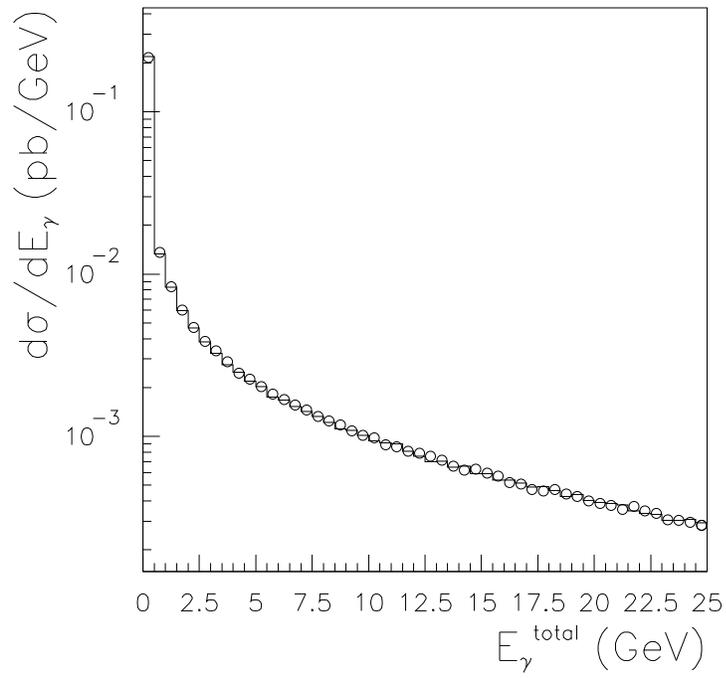}
}
\caption{
Differential cross section of the total energy of emitted photon(s)
obtained from the QEDPS (histogram) and from the SF(circle).
}
\label{fig-grace-singlw-1}
\end{figure}

\begin{figure}[htb]
\centerline{
\epsfysize=10cm
\epsfbox{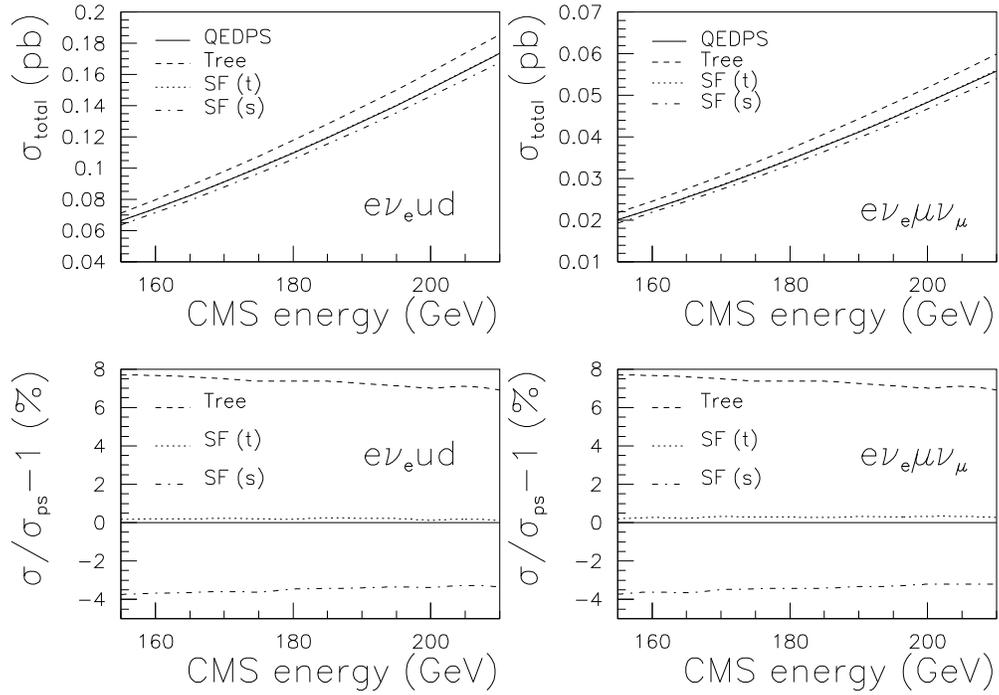}
}
\caption{
Total cross sections and those normalized by the QEDPS results
for $e\nu_e\bar{u}d$ and $e\nu_e\mu\nu_{\nu}$ processes 
without experimental cuts.
The SF(t) denotes the SF with proper energy scale and the SF(s) with the inappropriate
energy scale ($s$). 
Only $t$-channel diagrams(non-annihilation diagrams) are taken into 
account.
}
\label{fig-grace-singlw-2}
\end{figure}

For the total energy of the emitted photons both methods, SF and QEDPS, 
must give the same spectrum once the same energy scale is used. 
This is confirmed for the semi-leptonic process as shown in 
Fig.\ref{fig-grace-singlw-1} at the CM energy of 200 GeV.
We choose the same energy-evolution scale as SF described by 
Eq.(\ref{correction}).
Total cross sections as a function of the CM energies at LEP2 
with no cuts are shown in Fig.\ref{fig-grace-singlw-2}. 
The effect of the QED
radiative corrections on the total cross sections are obtained to be
7 to 8\% in the LEP2 energy region. 
If one used an inappropriate energy scale, say $s$, in SF, the ISR effect
is overestimated about 4\% as seen in Fig.\ref{fig-grace-singlw-2}.
The result of SF is consistent with the QEDPS around 0.2\% if
the proper energy scale is employed.
 
There is some ambiguity to choose the energy-evolution scale in the
leading-log order. For example the transverse-momentum squared or 
the invariant-mass squared of $u\bar d$ (or $\mu {\bar\nu_\mu}$) systems
could be a candidate of the energy scale.
It is found that even these energy scales are chosen in QEDPS
instead of $(p_+-(k_u + k_d))^2$, the total cross sections change by
only 0.6\%, which must be allocated to the theoretical uncertainty.

\begin{figure}[htb]
\centerline{
\epsfysize=15cm
\epsfbox{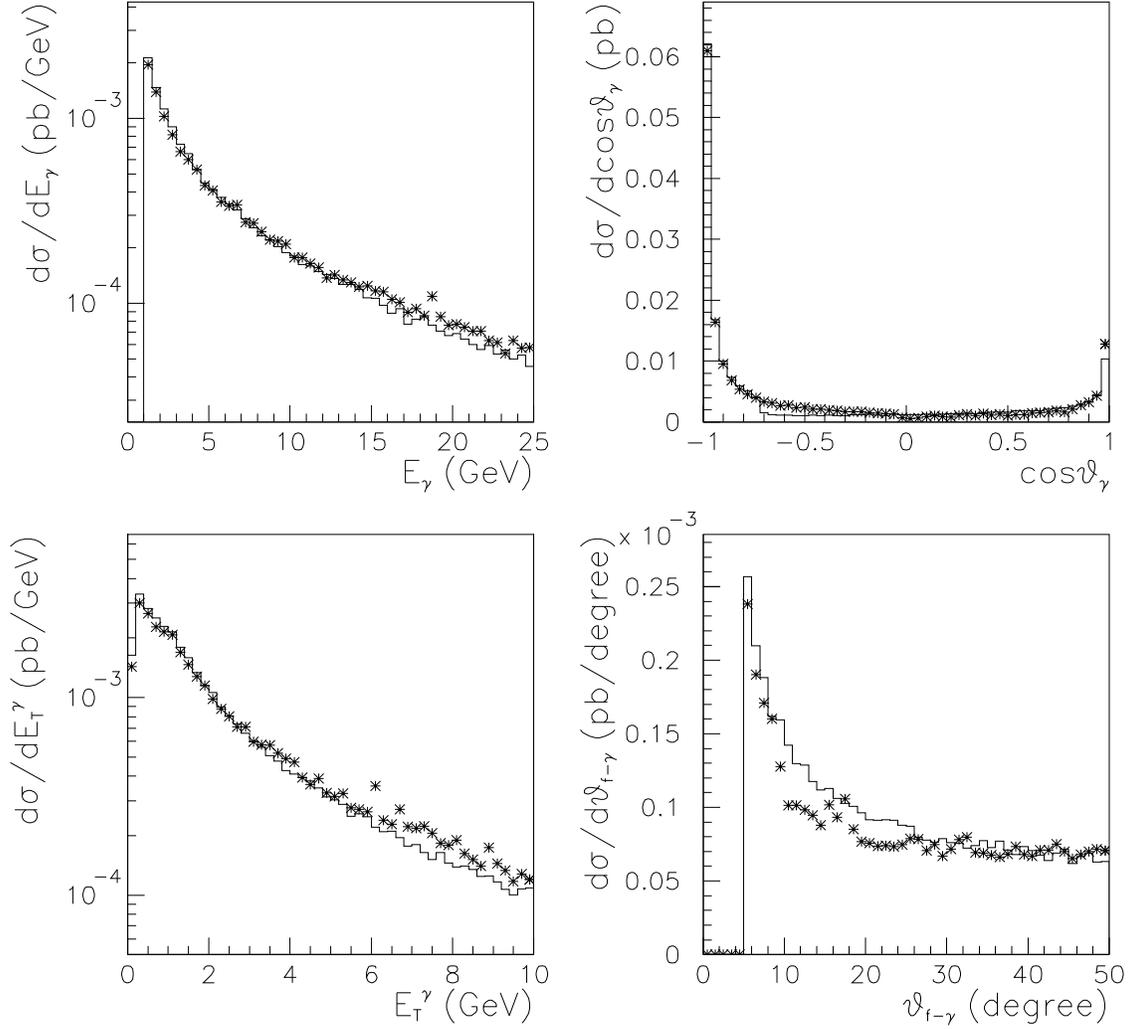}
}
\caption{
Differential cross sections of the hard photon;
Energy, transverse energy w.r.t. the beam axis,
cosine of the polar angle, and
opening angle between photon and
nearest charged-fermion.
A histogram shows the QEDPS result and stars from the matrix element
with soft-photon correction.
Only $t$-channel diagrams(non-annihilation diagrams) are taken into 
account.
}
\label{fig-grace-singlw-4}
\end{figure}

%% hard-photon emission
The energy and angular distributions of the hard photon from QEDPS
are compared with those obtained from the exact matrix elements. 
The cross sections of the radiative process
$e^- e^+ \rightarrow e^- {\bar \nu}_e u {\bar d} \gamma$
are calculated based on the exact amplitudes generated by ${\tt GRACE}$
and integrated numerically in five-body phase space using ${\tt BASES}$.
Again only the $t$-channel diagrams(non-annihilation diagrams) are taken 
into account. To compare the distributions the soft-photon corrections for
the radiative process must be included. For this purpose
QEDPS is implemented into the $e^- {\bar \nu}_e u {\bar d} \gamma$
calculations with a careful treatment to avoid a double-counting of
the radiation effect. The definition of the hard photon is
\begin{enumerate}
\item $E_\gamma>1$ GeV,
\item opening angle between the photon and the nearest final-state 
charged particles is greater than $5^\circ$.
\end{enumerate}
The distributions of the hard photons are in good agreement as seen
in Fig.\ref{fig-grace-singlw-4}.
The total cross section of the hard photon emission agrees each other 
in 2\% level. 
We also calculated the radiative cross-section without soft-photon
correction. If the soft-photon correction is not 
included in the radiative process, we find a 30\% overestimation of the
radiative cross-section with above experimental cuts.

\subsection{Cross sections with experimental cut}
\begin{figure}[htb]
\centerline{
\epsfysize=10cm
\epsfbox{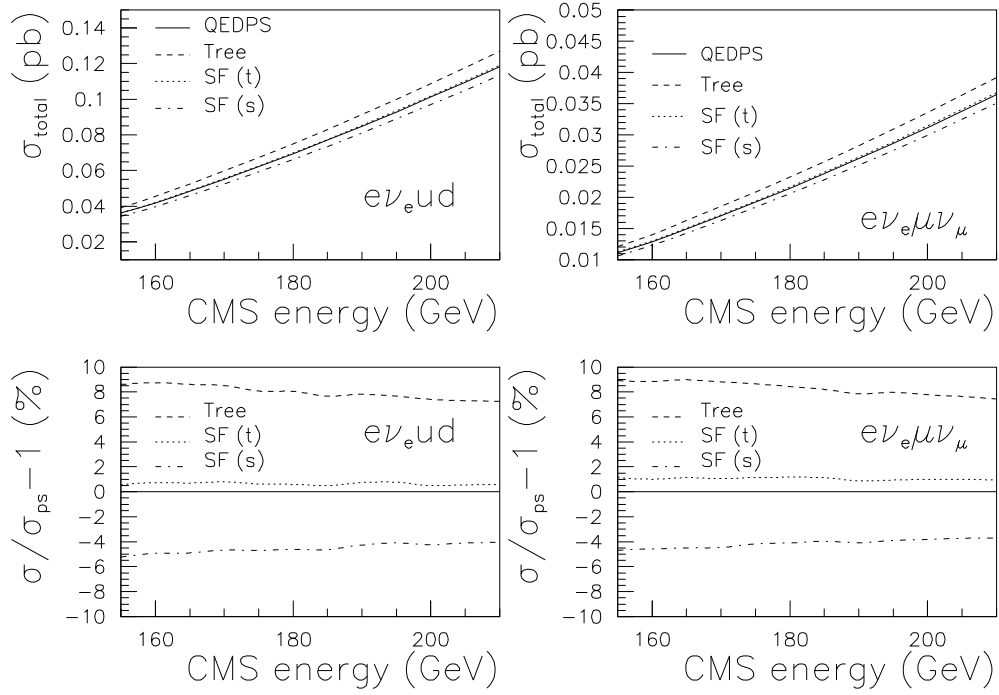}
}
\caption{
Total cross sections and those normalized by the QEDPS results
for $e\nu_e\bar{u}d$ and $e\nu_e\mu\nu_{\nu}$ processes 
with experimental cuts.
The SF(t) denotes the SF with proper energy scale and the SF(s) 
with the inappropriate
energy scale ($s$).
All diagrams are taken into account.
}
\label{fig-grace-singlw-3}
\end{figure}

It is also investigated how large the effects of the QED radiative 
corrections is for the single-$W$ production when realistic experimental 
conditions are imposed. The experimental cuts applied here are
$\theta_{e^-} < 5^\circ$,
$M_{q {\bar q}}>45~\hbox{GeV}$,
$E_{\mu}>20~\hbox{GeV}$.
For this study all the diagrams, not only the $t$-channel diagrams but
also the $s$-channel, are taken into account in the calculations.
The dominance of the signal from $t$-channel diagrams is 97\%(90\%)
for the hadronic(leptonic) decay of the $W$ boson, respectively.
Total cross sections as a function of the CM energies at LEP2 with
these cuts are shown in Fig.\ref{fig-grace-singlw-3}. 
The QED radiative corrections on the total cross sections are found to 
be 7 to 8\% in this LEP2 energy range. If one uses the inappropriate energy scale 
$s$ in SF, the ISR effect is overestimated by around 5\%, which is larger 
than those of the no-cut case. SF with the proper energy scale shows 
a deviation from QEDPS around 0.5\% for $e^- {\bar \nu}_e u {\bar d}$ 
process and 1.0\% for $e^- {\bar \nu}_e \mu^+ \nu_\mu$. 
The agreement between QEDPS and SF becomes worse than the no-cut 
case, because the finite transverse-momentum of the emitted 
photons by QEDPS changes the acceptance of the above electron-veto 
requirement. Hence for the realistic experimental conditions 
the finite transverse-momentum of emitted photons should not be ignored.

\begin{figure}[htb]
\centerline{
\epsfysize=10cm
\epsfbox{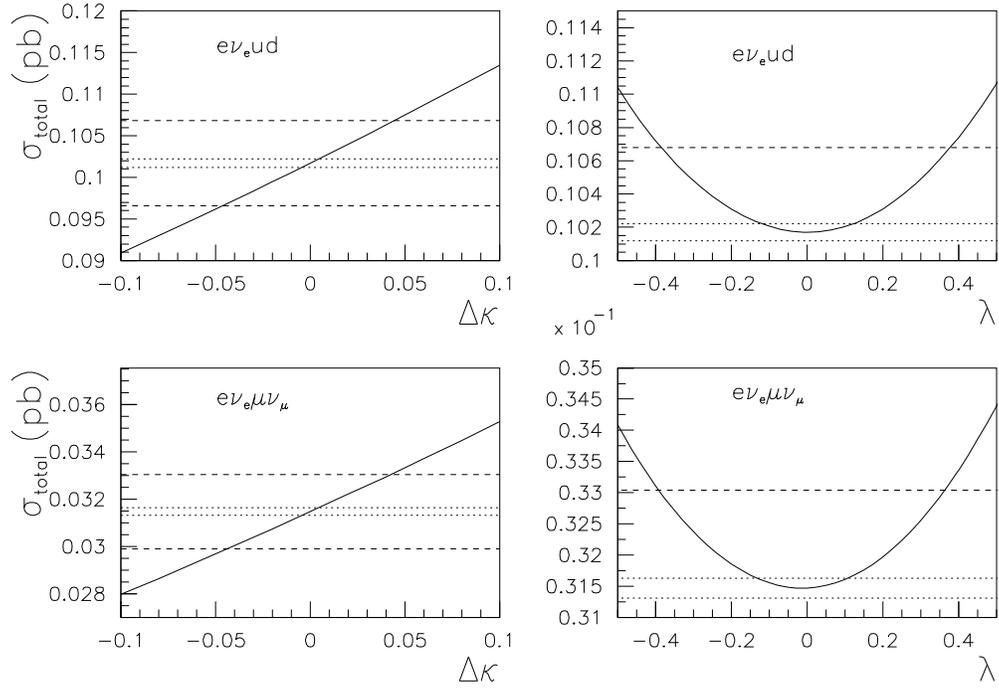}
}
\caption{
Total cross-sections of $e\nu_e\bar{u}d$ and $e\nu_e\mu\nu_{\nu}$ processes 
with experimental cuts as a function of the anomalous TGC.
Dashed (dotted) lines show $\pm5$\% ($\pm0.6$\%) cross section variation 
from its standard model value.
The cross sections are obtained at the CM-energy of 200 GeV with 
all diagrams.
}
\label{fig6}
\end{figure}

Finally a possible effect of the systematic error of the ISR effect on the
anomalous TGC measurements is investigated. 
If the inappropriate energy scale is used in
the ISR tools such as the CM-energy squared, the total cross sections
with the experimental cuts includes 5\% systematic error.
Even if one used one of proper energy scales, there is 0.6\% uncertainty
on the cross sections due to ambiguity of the energy scale selection. 
These uncertainty of the total cross sections limits the experimental
sensitivity of the anomalous TGC measurements. 
Total cross-sections of $e\nu_eud$ and $e\nu_e\mu\nu_{\nu}$ processes 
with experimental cuts as a function of the anomalous TGC is summarized
in Fig.\ref{fig6}.
The cross sections are obtained at the CM-energy of 200 GeV with 
all diagrams.
A bound shown in dashed lines shows $\pm5$\% cross section variation 
from its standard model value. If one used the inappropriate energy scale,
it affects $\pm 0.05$ on $\Delta \kappa$ measurement and
$\pm 0.4$ on $\lambda$ measurement.
The ambiguity of the energy-scale selection gives the systematic error
of less than 0.01 on $\Delta \kappa$ and 0.1 on $\lambda$.

\section{Conclusions}
The method to apply the QED radiative correction to the non-annihilation
process was established. The conventional method, SF with the energy scale 
$s$ gave about 4\% overestimation in the LEP2 energies. The uncertainty 
due to the energy-scale determination was estimated to be about 0.6\%. 
This uncertainty may affect the anomalous TGC measurements 
less than 0.01 on $\Delta \kappa$ and 0.1 on $\lambda$.
If one wants to look at the hard photon spectrum, the soft-photon 
correction to these radiative processes are needed.

In this report we have treated two processes 
$e^- e^+ \rightarrow e^- {\bar \nu}_e u {\bar d}$ and 
$e^- e^+ \rightarrow e^- {\bar \nu}_e \mu^+ \nu_\mu$.
The CP conjugate processes 
$e^- e^+ \rightarrow e^+ \nu_e {\bar u} d$ and 
$e^- e^+ \rightarrow e^+ \nu_e \mu^- {\bar \nu}_\mu$ give the
same results and the other channels
$e^- e^+ \rightarrow e \nu_e c s$ and 
$e^- e^+ \rightarrow e \nu_e \tau \nu_\tau$ will show slightly
different results due to their masses.
On the other hand the self CP-conjugate process 
$e^- e^+ \rightarrow e^- {\bar \nu}_e e^+ \nu_e$ has an additional 
complexity, because two energy scales $t_- = (p_{e^-}-q_{e^-})^2$ 
and ${\tilde t}_- = (p_{e^-}-q_{e^-}-q_{{\bar \nu}_e})^2$ 
can appear simultaneously. 

\vskip 1cm
Authors would like to thank members of the four-fermion
working group of the LEP2 Monte Calro Workshop at CERN, in particular
A.~Ballesterero, G.~Montagna, F.~Piccinini, G.~Passarino for useful
discussions.

This work was supported in part by the Ministry of Education, Science
and Culture under the Grant-in-Aid No. 11206203 and 11440083.

%%%%%%%%%%%%%%%%%%%%%%%%%% ref %%%%%%%%%%%%%%%%%%%%%%%%%%%%%%%%%%%%%%%

\newpage
%%%%%%%%%%%%%%%%%%%%%%%% fig 1 %%%%%%%%%%%%%%%%%%%%%%%%%%%%%%%%%%%%%%%%
%%%%%%%%%%%%%%%%%%%%%%%% fig 2 %%%%%%%%%%%%%%%%%%%%%%%%%%%%%%%%%%%%%%%%
%%%%%%%%%%%%%%%%%%%%%%%% fig 3 %%%%%%%%%%%%%%%%%%%%%%%%%%%%%%%%%%%%%%%%
%%%%%%%%%%%%%%%%%%%%%%%% fig 4 %%%%%%%%%%%%%%%%%%%%%%%%%%%%%%%%%%%%%%%%
%%%%%%%%%%%%%%%%%%%%%%%% fig 5 %%%%%%%%%%%%%%%%%%%%%%%%%%%%%%%%%%%%%%%%
%%%%%%%%%%%%%%%%%%%%%%%% fig 6 %%%%%%%%%%%%%%%%%%%%%%%%%%%%%%%%%%%%%%%%
%%%%%%%%%%%%%%%%%%%%%%%%%%%%%%%%%%%%%%%%%%%%%%%%%%%%%%%%%%%%%%%%%%%%%%%
\end{document}